\documentclass[manuscript,screen]{acmart}
\usepackage{graphicx}
\usepackage{multirow}
\usepackage{subcaption}
\usepackage{booktabs}
\usepackage{titlesec}
\DeclareUnicodeCharacter{2212}{-}
\AtBeginDocument{%
  \providecommand\BibTeX{{%
    \normalfont B\kern-0.5em{\scshape i\kern-0.25em b}\kern-0.8em\TeX}}}

\setcopyright{acmlicensed}
\copyrightyear{2024}
\acmYear{2024}
\acmDOI{XXXXXXX.XXXXXXX}

\begin{document}

\title{Parallel Stacked Aggregated Network for Voice Authentication in IoT-Enabled Smart Devices}

\author{Awais Khan}
\email{awaiskhan@oakland.edu}
\orcid{0000-0003-2497-7687}
\affiliation{%
  \institution{School of Engineering and Computer Science, Oakland University}
  \city{Rochester}
  \state{Michigan}
  \country{USA}
  \postcode{48309-4479}
}

\author{Ijaz Ul Haq}
\email{ijazulhaq@oakland.edu}
\affiliation{%
  \institution{College of Innovation and Technology, University of Michigan-Flint}
  \city{Flint}
  \state{Michigan}
  \country{USA}
  \postcode{48502-1950}}

\author{Khalid Mahmood Malik}
\email{mahmood@oakland.edu}
\email{drmalik@umich.edu}
\affiliation{%
 \institution{Oakland University, University of Michigan-Flint}
 \city{Flint}
 \state{Michigan}
 \country{USA}
 \postcode{48502-1950}
 }

\renewcommand{\shortauthors}{Khan, et al.}

\begin{abstract}
Voice authentication on IoT-enabled smart devices has gained prominence in recent years due to increasing concerns over user privacy and security. The current authentication systems are vulnerable to different voice-spoofing attacks (e.g., replay, voice cloning, and audio deepfakes) that mimic legitimate voices to deceive authentication systems and enable fraudulent activities (e.g., impersonation, unauthorized access, financial fraud, etc.). Existing solutions are often designed to tackle a single type of attack, leading to compromised performance against unseen attacks. On the other hand, existing unified voice anti-spoofing solutions, not designed specifically for IoT, possess complex architectures and thus cannot be deployed on IoT-enabled smart devices. Additionally, most of these unified solutions exhibit significant performance issues, including higher equal error rates or lower accuracy for specific attacks. To overcome these issues, we present the parallel stacked aggregation network (PSA-Net), a lightweight framework designed as an anti-spoofing defense system for voice-controlled smart IoT devices. The PSA-Net processes raw audios directly and eliminates the need for dataset-dependent handcrafted features or pre-computed spectrograms. Furthermore, PSA-Net employs a split-transform-aggregate approach, which involves the segmentation of utterances, the extraction of intrinsic differentiable embeddings through convolutions, and the aggregation of them to distinguish legitimate from spoofed audios. In contrast to existing deep Resnet-oriented solutions, we incorporate cardinality as an additional dimension in our network, which enhances the PSA-Net ability to generalize across diverse attacks. The results show that the PSA-Net achieves more consistent performance for different attacks that exist in current anti-spoofing solutions. Moreover, PSA-Net outperforms most dedicated attack detection systems and offers a single, secure solution for voice-controlled smart IoT devices.

\end{abstract}


\received{April 2024}

\maketitle

\section{Introduction}
Internet of Things (IoT)-enabled voice-controlled devices are gaining widespread adoption due to the convenience they offer users in remotely controlling domestic appliances (e.g., smart doors, surveillance, or home appliances) using voice assistants. Voice assistants (like Apple Siri, Google Home, and Amazon Alexa, etc.) enable users to make calls, send emails, control smart devices, and even perform various banking tasks. This ease of use has accelerated rapid adoption, with forecasts exceeding \$31 billion in the market share~\cite{park2020effective} and 157 million users predicted in the United States by 2026~\cite{statista_2023_voice_assistant}. Beyond convenience, voice assistants offer potential safety benefits. For instance, while driving, users can control features like music playlists without taking their hands off the wheel, which minimizes distractions and enhances road safety. Similarly, voice interaction promotes social distancing by eliminating the need for traditional touch-based interactions (e.g., in COVID-19). Nevertheless, despite their widespread adoption and promising features, the growing reliance on voice technology comes with critical security vulnerabilities. For example, due to their reliance on voice authentication, these systems are susceptible to voice-spoofing attacks, where malicious actors mimic authorized voices to gain access, perform unauthorized actions, or even steal sensitive information. From fraudulent transactions to compromised security, the consequences of successful spoofing attacks can be severe. As voice control continues to evolve our lives, addressing these security challenges is essential to ensuring user trust and the ethical development of the technology.

\begin{figure}[!b]
        \centering
        \includegraphics[width=15cm]{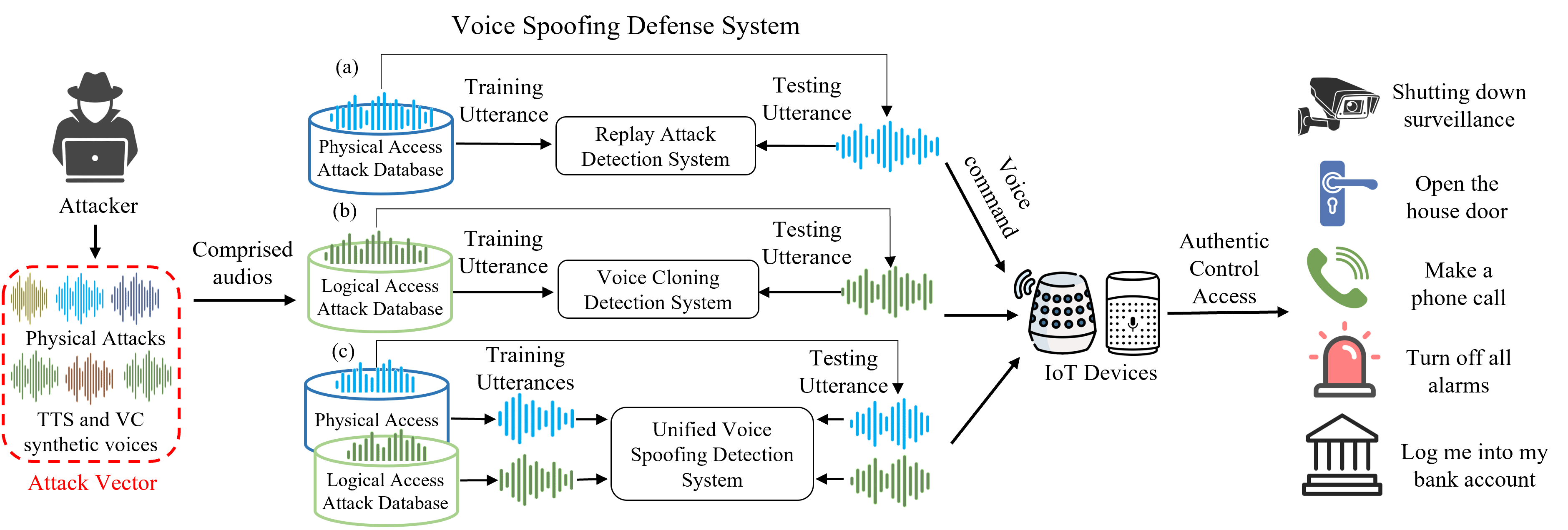}
        \caption{The workflow of voice-spoofing attacks and the corresponding defense mechanisms safeguarding IoT-enabled smart devices. (a) Replay Attack Detection: a standard system trained on recorded physical access attacks (replays) effectively detects attempted replayed commands. (b) Voice Cloning Detection: A dedicated system trained on text-to-speech (TTS) and voice conversion data to identify spoofing from logical access. (c) Unified Detection: A single system trained to detect all types of spoofing attempts. This system ensures only verified commands reach the IoT device and execution of the commands within authorized access.}
        \label{fig:UnifiedVsSingular}
\end{figure} 

Voice-spoofing attacks, broadly categorized into physical access (PA) and logical access (LA) attacks, can compromise sensitive information stored on various IoT devices. These devices (i.e., smartphones, smartwatches, and smart speakers) store a variety of user data, including emails containing sensitive information (e.g., social security numbers, credit cards, pictures, and contacts) that might be accessed and used to threaten the user. The most representative spoofing attacks are replay and voice cloning attacks. Specifically, replay attacks involve replaying previously recorded voice commands, while voice cloning utilizes modern AI algorithms to generate synthetic voices. Both types of attacks aim to deceive the targeted devices into engaging in harmful activities. Consequently, detecting these attacks is crucial for the continued adoption and safe use of IoT enabled voice-controlled devices.

In order to secure IoT enabled voice-controlled devices, various strategies have been developed to combat voice-spoofing, as illustrated in Figure~\ref{fig:UnifiedVsSingular}. In particular, the current voice-spoofing defense system (VSD) in IoT enabled devices are designed for replay and voice cloning attacks. However, prior studies~\cite{das2020assessing, khan2023battling} show that different attacks cause distinct irregularities in the audio. For instance, voice-cloned speeches lacked ambient elements and natural human pauses, while in replays, subtle audio variations like micro-phonic inconsistencies are induced due to variant microphones and recording devices. These distinctive characteristics make existing VSDs trained for one attack ineffective for others. Thus, there is a need for a more versatile, unified solution that can effectively address different types of voice-spoofing within a single system simultaneously.

Due to the contrastive nature of attacks, most voice-spoofing countermeasures focus on a single type of spoofing attack. However, in recent years, the research community has tended to design unified solutions for all types of spoofing attacks. Despite some progress, existing VSDs show significant performance differences across different attacks~\cite{8272715}. For instance, a study~\cite{lai2019assert} reported an equal error rate (EER) performance of $0.59\%$ for replay attacks, but this EER raises to $6.70\%$ for voice cloning attacks, highlighting a significant performance drop. Similar performance gaps are observed in traditional VSDs like STC~\cite{lavrentyeva2019stc}, BUT-Omilia~\cite{zeinali2019detecting}, MFMT~\cite{li2019anti}, and in recent solutions like Convnext~\cite{ma2022convnext}, SASV~\cite{aljasem2021secure}, and others~\cite{sinha2024improving,liu2024multi,huang2024self}. These performance deficiencies indicate a broader security challenge for voice-controlled IoT-enabled devices. On the one hand, these current VSDs are not well generalized and often struggle to address emerging attacking variations like partial spoof~\cite{zhang2021initial}. On the other hand, most of the VSDs rely on resource-intensive computations (e.g., spectrograms and handcrafted features), which may not be suitable for IoT enabled resource-constrained devices. 

Consequently, from the literature we observed that the existing unified VSDs suffer from two significant limitations: firstly, the lack of generalization and equally good performance for unknown and unseen attacks. Secondly, the structure of existing solutions is computationally complex, which makes them unsuitable for IoT-enabled, resource-constrained devices. These limitations motivate us to analyze and address key research concerns regarding the equally optimal detection capabilities of VSDs and the analysis of optimal inputs for DNN models in resource-constrained devices. In response, in this paper, we propose the Parallel-Stack Aggregated Network (PSA-Net), a novel lightweight architecture for unified spoofing detection. Drawing inspiration from the success of ResNeXt models in computer vision~\cite{xie2017aggregated}, PSA-Net utilizes a wider, multi-branch network architecture with an identical topology. Leveraging the efficient "split-transform-merge" approach, PSA-Net seamlessly extracts and combines acoustic cues through its repeating residual architecture, resulting in robust voice embeddings for classifying spoof and genuine speech. Notably, PSA-Net operates directly on raw speech signals and bypass computationally expensive transformations (e.g., spectrograms) and handcrafted features which enhance its lightweight nature. Moreover, PSA-Net addresses performance gaps against diverse attacks by incorporating data augmentation to balance the training dataset and prevent overfitting by employing spatial dropout to mitigate the vanishing gradient problem. This combination of strategies leads to improved generalization and robustness of our proposed PSA-Net in comparison to the existing solutions. The main contributions of this work are summarized as follows:

\begin{enumerate}
    \item We introduce a squeeze- and excitation-based parallel-stacked aggregation network (PSA-Net) for detecting diverse voice-spoofing attacks within a single framework. The network uses raw audio directly, minimizing the computational complexities associated with spectrograms or hand-crafted features making it compatible for voice-controlled smart IoT devices.
    \item We employed a multi-branch aggregation architecture to effectively distinguishing real from artificially generated or replay fake voice commands. The proposed solution notably mitigates the bias observed in state-of-the-art unified solutions towards a particular attack class.
    \item We examine the efficacy of several residual and aggregated residual networks with squeeze and excitation (SE) blocks and provide a detailed ablation study for unified voice-spoofing detection. To the best of our knowledge, this is a pioneering work utilizing SE in a stacked aggregated network for unified spoofing attack detection.
    \item We performed a detailed experimental analysis to test the effectiveness of the proposed solution with four distinct datasets (ASVspoof-2019, ASVspoof-2021, Partial spoof, and VSDC) against several attack-specific and unified solutions. Moreover, we also performed experimental analysis over resource constrained devices that shows the compatibility of the proposed solutions with IoT-edge devices.
\end{enumerate}{}{}

The rest of the paper is organized as follows: Section~\ref{existing work} provides the overview of existing work and limitations of the unified solutions, Section~\ref{proposed work} presents the detail architecture of proposed work. Section~\ref{Experiment} outlines the experimental setup, and Section~\ref{results} provides the details about the experimental results and discussion along with the ablation study. Finally, Section~\ref{conclusion} presents the conclusion of this work.

\section{Existing Work}\label{existing work} 
The existing VSDs for replay or voice cloning attacks can be classified into two main categories: hand-crafted countermeasures with frame-level classifiers such as GMM and i-vectors~\cite{khan2022voice}, and deep learning-based end-to-end solutions~\cite{li2021replay,huang2024self}. Each category can further be categorized into ~\ref{SS} Standalone dedicated VSDs designed for specific attack types and ~\ref{US} Unified solutions encompassing all types of voice-spoofing attacks.

\subsection{Standalone Dedicated Voice-spoofing Detection}
\label{SS}
Most research on voice-spoofing attacks is primarily focused on developing attack-specific VSDs using traditional handcrafted features and frame-level classifiers~\cite{sahidullah2015comparison, sriskandaraja2016front, albawi2017understanding}. These methods first process the input audio for feature extraction (e.g., MFCC \cite{zheng2001comparison}, CQCC \cite{todisco2016new}, and later perform classification using a backend classifier (e.g., GMM ~\cite{sahidullah2015comparison}, i-vector~\cite{rahmeni2020speech}, etc.). The features intended to detect the underlying artifacts of spoofing attacks can be categorized by the focused regions (e.g., spectral \cite{paul2017spectral}, cepstral~\cite{todisco2016new}, etc.) and the filters (e.g., linear, triangular, etc.) used to perform the extraction of the features. For instance, magnitude spectrum-based \cite{kamble2020amplitude}, phase spectrum-based \cite{yang2021modified}, energy-based \cite{gupta2023replay}, modulation-based \cite{kamble2020amplitude}, and others \cite{aljasem2021secure, javed2022voice, huang2020audio}. Among these, linear frequency and cepstral coefficients (LFCC) have shown greater detection performance in the prior ASVspoof challenges~\cite{sahidullah2015comparison}. Although most of these features are used along with traditional backend classifiers (e.g., GMM~\cite{sahidullah2015comparison,rahmeni2020speech} and SVM~\cite{khan2022voice, khan2022toward}), some recent studies applied deep classifiers to perform the classification of voice samples \cite{huang2020audio, khan2023spotnet}. Further, several studies combine multiple types of features to obtain unique spoofing attributes~\cite{omeroglu2022multi}. While these techniques perform well against trained-on attacks, their feature design processes require significant domain expertise, and they may suffer from limited generalizability against new types of attacks like partial spoofs \cite{zhang2021initial}.

Following the rise of deep neural networks, researchers explored integrating handcrafted features with various architectures, like CNN~\cite{albawi2017understanding}, ResNet~\cite{li2021replay}, and attention-based methods~\cite{yu2018deep}. This led to VSDs that combined the strengths of DNNs with handcrafted features, resulting in capturing more discriminative featrues~\cite{jung2022aasist, zhang122021effect, tak2021end, omeroglu2022multi, ma2021rw}. While these methods outperformed traditional solutions, i.e., \cite{ sahidullah2015comparison, yang2021modified, aljasem2021secure}, they often focus on a single type of spoofing attack, limiting their effectiveness against the diverse nature of attacks.

\subsection{Unified Voice-spoofing Attack Detection}
\label{US}
In recent years, few studies have attempted to create unified VSD systems aiming to detect a spectrum of spoofing attacks within a single framework. For instance, in \cite{lavrentyeva2019stc}, the author presents a unified VSD system based on different front-end features. However, although the authors found that the LFCC-based system performed better than the others, it failed to detect PA attacks effectively. Further, the EER of the LFCCs was reported to be higher when tested against replay attacks. Similarly, Li et al.~\cite{li2019anti} employed multi-task learning with multi-feature integration (MFCC \cite{zheng2001comparison}, CQCC \cite{todisco2016new}, and Filter Bank) to detect both replay and voice cloning attacks. Although this combination of cepstral features performed well for replay attack detection (EER of $0.96\%$), it struggled with speech synthetic and voice conversion attacks. In contrast to the cepstral features, a combination of ternary features, named sm-ALTP \cite{aljasem2021secure}, was presented to perform unified spoofing detection. Even after suppressing the performance of cepstral features with aggregation-based ensemble, the ternary features struggled against synthetic-based attacks \cite{aljasem2021secure}. In another study, Zeinali et al. \cite{zeinali2019detecting} presented an VSD system that merged two VGG networks trained on single- and two-feature sets. While it performed reasonably well with power spectrogram and CQT features, it encountered higher EERs in synthetic-based attacks. Further, in the ASSERTS system \cite{lai2019assert}, CQCC features and a squeeze-and-excitation-based residual network were used to identify all types of spoofing attacks, such as replay-based, synthetic-based, and voice conversion. Although the model excelled in replay detection with a low $0.59\%$ EER, it struggled to identify synthetic-based attacks, resulting in an increased EER of $6.70\%$ during evaluation.

Recently, Suthokumar et al.~\cite{{8913369}} employed the feature fusion of three handcrafted features and a deep neural network to develop a unified solution. Similarly, Das et al.~\cite{das2020assessing} analyze the generalization of the VSDs by using three handcrafted features within a similar approach. However, these methods mixed the LA and PA attack data to train their solution and thus lacked sufficient data diversity, potentially limiting their model performance compared to other existing models and recently developed attacks. In contrast to feature fusion, Monteiro et al.~\cite{monteiro2020generalized} proposed an ensemble-based approach with specialized models for each attack type and later ensemble them to combine their scores. Although, their solution achieves comparatively good accuracy but not compatible with IoT-enabled devices. 
The inconsistent performance of both standalone and unified VSD systems against state-of-the-art (SOTA) spoofing attacks raises serious concerns about the security of IoT enabled voice control systems. Moreover, the performance variation highlights the urgency for a robust VSD system capable of detecting SOTA spoofing attacks with equally good performance.

\begin{figure*}[!t]
        \centering
        \includegraphics[width=8cm, height= 12cm]{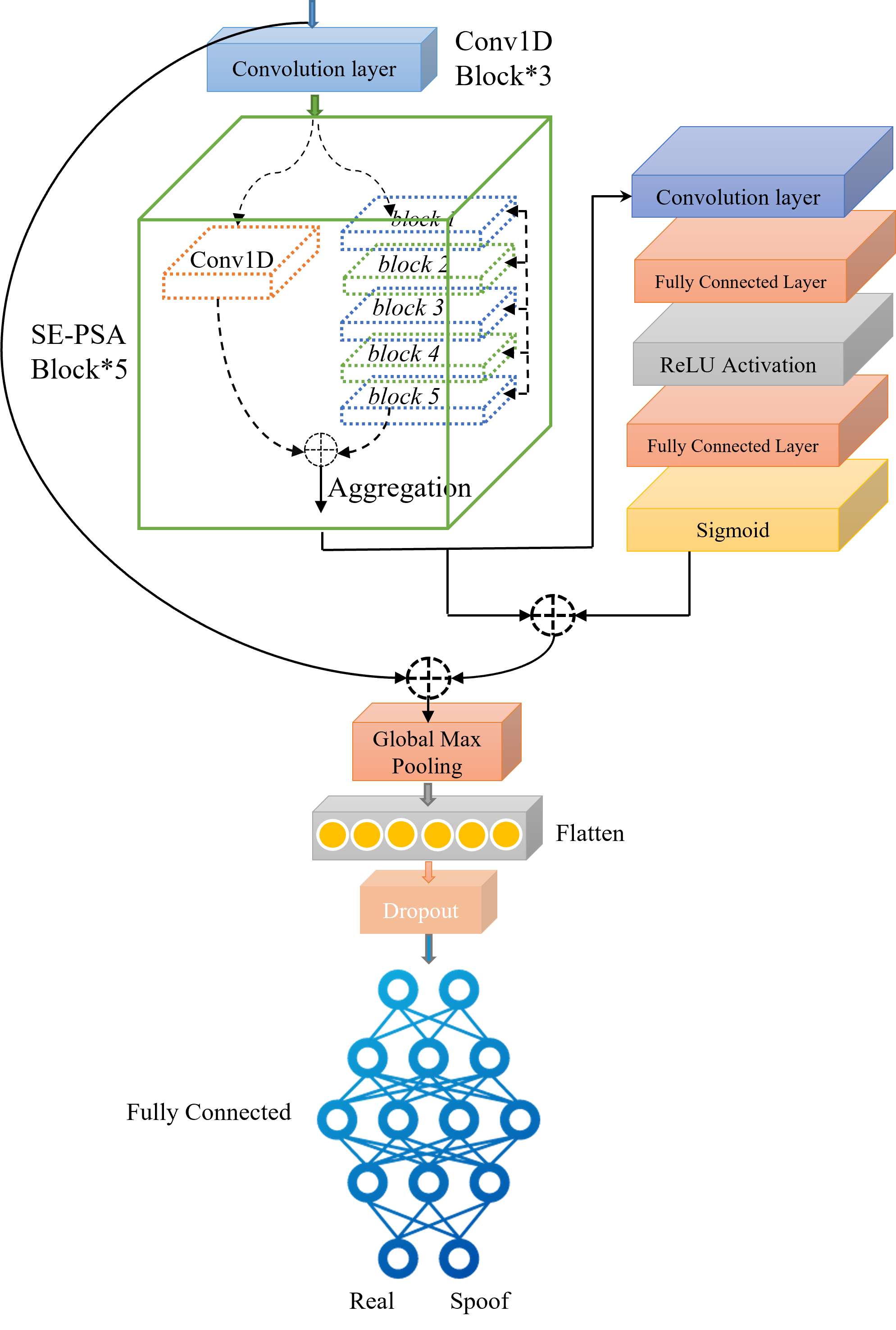}
        \caption{The internal architecture of the proposed parallel-stacked aggregated network (PSA-Net). Raw audio passes through three blocks of convolutional layers before it is split into various cardinalities. Residual connections are used between the layers, followed by global max pooling. The architectural design includes three Conv1D and five SE-PSA blocks before reaching the fully connected layer for real vs. spoof classification.}\vspace{-10pt}
        \label{fig:overallFramework}
\end{figure*} 
\section{Proposed Framework}\label{proposed work}  
This section presents technical details of PSA-Net, a novel architecture for unified voice-spoofing detection. PSA-Net is divided into two distinct parts; 1) data preparation that encompasses both preprocessing and augmentation techniques to prepare the raw voice data for the network (Sections~\ref{pp} and ~\ref{DA}). 2) extracting convolved audio representations and fine-grained embeddings using parallel stacking and aggregation strategy to classify the audio signal into real and fake. 

\subsection{Overview}
The proposed PSA-Net follows a topology identical to ResNeXt's intra-architecture \cite{xie2017aggregated} having ability to reduce the risk of hyper-parameter over-adaptation, which is currently missing in existing ResNet models. Aggregating wider networks instead of deeper ones has already proven effective performance in computer vision~\cite{wu2019wider}, while their application in audio classification or spoofing detection remains unexplored. This work aims to fill this gap by proposing PSA-Net, which leverages the Split-Transform-Merge (STM) strategy and squeeze-and-excitation approaches for unified voice-spoofing detection. Additionally, PSA-Net utilizes a stacking-based classifier to effectively discern spoofing artifacts in both genuine and spoofed speech samples. Notably, the STM approach achieves optimal performance with reduced computational complexity, making PSA-Net suitable for IoT based resource-constrained devices.

\subsection{Parallel Stacking and Aggregation}  
The PSA-Net architecture consists of two main components: convolution blocks (Conv1D) for extracting initial convolved audio embeddings from preprocessed voice signals (see Sections ~\ref{DA} and~\ref{pp}) and SE-PSA blocks, that analyze the extracted initial convovled embeddings to capture fine-grained features embeddings for classifying spoofed and genuine speech samples. Both components follow a parallel topology, where each block processes the splitted input independently before aggregating the results as described in a subsection below. Figure \ref{fig:overallFramework} provides a visual representation of both the overall and internal architecture of the PSA network. In particular, convolution blocks utilize three layers with varying filter sizes and strides to extract initial embeddings. While SE-PSA blocks employ a repeating residual strategy with a split-transform-merge strategy, where each block comprises four branches with group convolutions to generate robust feature representations. The blocks progressively increase in width as the processing progresses. The final stage utilizes a fully connected layer to classify the speech sample as genuine or spoofed. Detailed specifications of filter sizes, strides, and activation functions are provided in Table \ref{tab:PSA}.

\begin{table}[t]
\caption{Detailed architecture of the PSA-Net. Adaptive average pooling and global average pooling are selected as pooling layers, and the output size is set to (1,1), which results in each channel having exactly one output to feed to the fully connected layers.}\vspace{-8pt}
\label{tab:PSA}
\centering
\begin{tabular}{cccc }
\toprule
  Layer & Output Size & SE-PSA  & Channel  \\
\midrule
 \multirow{2}{1cm}{Conv1D} & \multirow{2}{2cm}{$T \times F $ }& \multirow{1}{2.2cm}{$7 \times 7$ , 64, stride 2}   & \multirow{2}{0.4cm}{16} \\ 
           &           &        $max pool, stride 2 $    &   \\ \hline
 \multirow{2}{1cm}{SE-PSA} &\multirow{2}{2cm}{$T \times F$ } & \multirow{1}{2.2cm}{$ \left[ 3\times 3,32\right] \times 2 $}  & \multirow{2}{0.4cm}{32}  \\
           &   & $\left[ 3\times 3,32 \right],  C=4 $  &    \\ \hline
\multirow{2}{1cm}{SE-PSA} &\multirow{2}{2cm}{$T/2 \times F/2$} & \multirow{1}{2.2cm}{$ \left[ 3\times 3,64\right] \times 2 $}  & \multirow{2}{0.4cm}{64} \\
           &   & $\left[ 3 \times 3,64 \right], C=4 $  &     \\ \hline
 \multirow{2}{1cm}{SE-PSA} & \multirow{2}{2cm}{$T/4 \times F/4$} & \multirow{1}{2.2cm}{$ \left[ 3\times 3,128\right] \times 2 $} & \multirow{2}{0.6cm}{128} \\
     &  & $\left[ 3 \times 3,128 \right], C=4$ &    \\ \hline
\multirow{2}{1cm}{SE-PSA} & \multirow{2}{2cm}{$T/4 \times F/4$}&\multirow{1}{2.2cm}{$ \left[ 3\times 3,256\right] \times 2 $}& \multirow{2}{0.6cm}{256} \\
      &   & $\left[ 3 \times 3,256 \right], C=4 $  &    \\ \hline
\multirow{2}{1cm}{SE-PSA} & \multirow{2}{2cm}{$T/8 \times F/8$ }& \multirow{1}{2.2cm}{$\left[ 3\times 3,512\right] \times 2 $} & \multirow{2}{0.6cm}{512} \\
        &           & $\left[ 3\times 3,512 \right], C=4  $  &   \\ \hline
$Output$& $1 \times 1$ &  $Global Max Pool$, 1000-d fc, sigmoid  & \multirow{2}{0.6cm}{$--$ }\\ 
\bottomrule    
\end{tabular}
 \end{table}

\subsubsection{Convolutional Audio Representation:}
In the initial stage, we feed the input audio signal denoted as $F[n]$, containing $N$ samples with spectral and temporal information, into a three-layer convolutional block. This block utilize various filters of size $c1, c2,$ and $c3$, with corresponding kernel sizes and strides of $k1, k2, k3$, respectively where the padding is kept uniform. For optimal performance, a pre-activation convolutional approach is employed, incorporating batch normalization, softmax activation, and a final convolution layer. This generates the covolutional audio feature map representing the speech signal with convolved representation. Subsequently, max pooling is used to further extract the enhanced audio embedding as $E^{st}_c = {e_1, e_2, e_3, \ldots, e_n}$, where $E^{st}_c$ represents the convolutional audio embeddings and $e$ denotes the embedding values used to extract the fine-grained feature embedding.

\subsubsection{Fine-Grained Feature Embedding:}
In the second stage, $ E^{st}_c$ is passed to the SE-PSA block to extract fine-grained representations denoted as $ F^{fg}_r = {f^{g}_1, f^{g}_2, f^{g}_3, \ldots, f^{g}_n}$ for the classification of genuine and spoofed speech samples. The core architecture of the PSA block adheres to the same two principles governing the ResNeXt architecture: Firstly, when creating spatial maps of identical size, blocks share identical hyper-parameters (width and filter sizes). Secondly, the width of the blocks is increased by a factor of $2$, leading to an increase in width each time the spatial map is down-sampled. This approach ensures efficient extraction of robust features for accurate classification. The architectural strategy of SE-PSA blocks (Split-Transformation and Merging) is presented in the next section below.

\subsubsection{Split-Transformation and Merging:}
\label{STM}
\begin{figure}[!t]
        \centering
        \includegraphics[width=8cm]{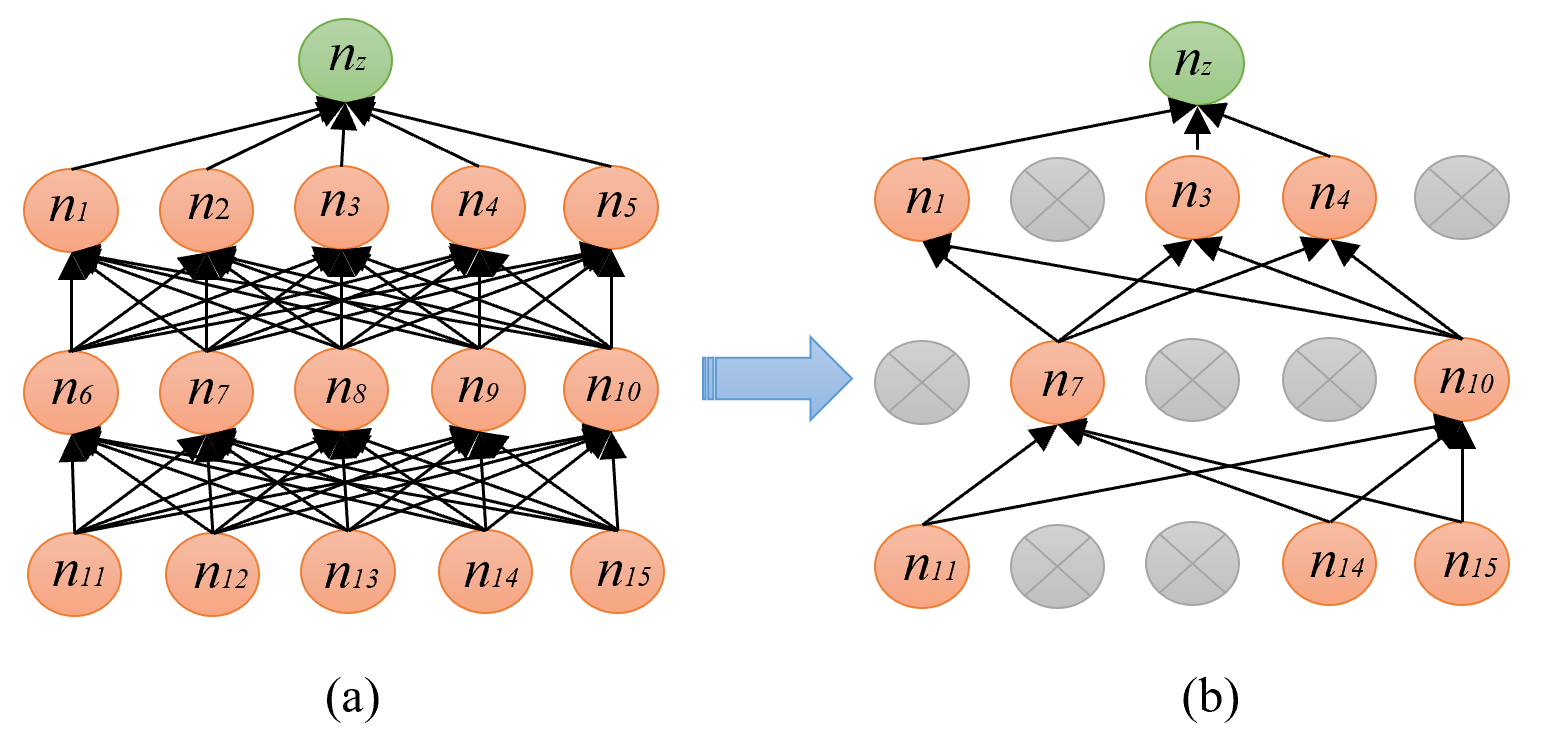}
        \caption{The internal architectural for addressing the vanishing gradient via Spatial dropout. (a) A standard DNN, with processing and activation of all neurons, without any selection or drop. (b) A standard neural network with Spatial dropout, which results in the selection of required neurons with more relevant embeddings as mentioned in \cite{lee2020revisiting}.} \vspace{-5pt}
        \label{fig:spatialdropout}
\end{figure}

\begin{figure}[!b]
        \centering
        \includegraphics[width=8cm, height=5.3cm]{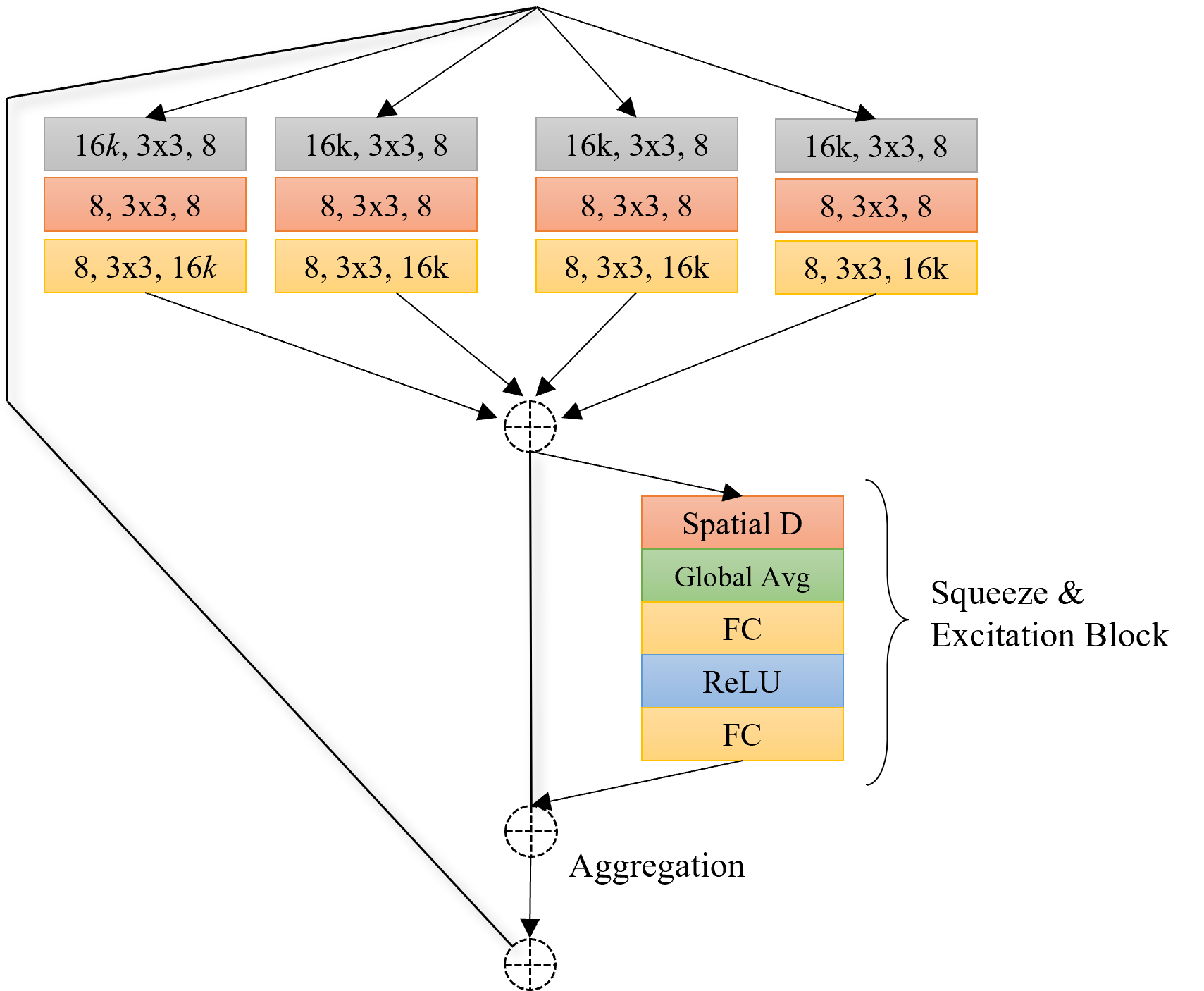}
        \caption{Aggregated Feature Map extraction with the Squeeze and Excitation Block. The SE block include the spatial dropout applied before every global average layer of the each SE-PSA block.} \vspace{-5pt}
        \label{fig:Squeeze&Excitation}
\end{figure}

PSA-Net leverages the "Split-Transform-Merge" approach within its SE-PSA blocks to extract fine-grained features crucial for spoof detection. As depicted in Figure \ref{fig:Intra-architecture}, each block combines the outputs $ F^{fg}_r$ of multiple parallel neural paths using concatenation. These paths, defined by a cardinality ${C}$ parameter, add width rather than depth to the network, allowing for robust feature representations. Subsequently, the obtained representations are passed to the next SE-PSA block. This process is iterated $M$ times to derive an adaptive feature representation for both spoofed and legitimate speech samples. The value of cardinality ${C}$ determines the size of the transformation set $T = {t_1, t_2, \ldots, t_n}$. The same transformations $T$ are applied $M$ times, and the cumulative gain is aggregated as shown in the equations below:
\begin{figure*}[!t]
        \centering
        \includegraphics[width=12cm, height= 10cm]{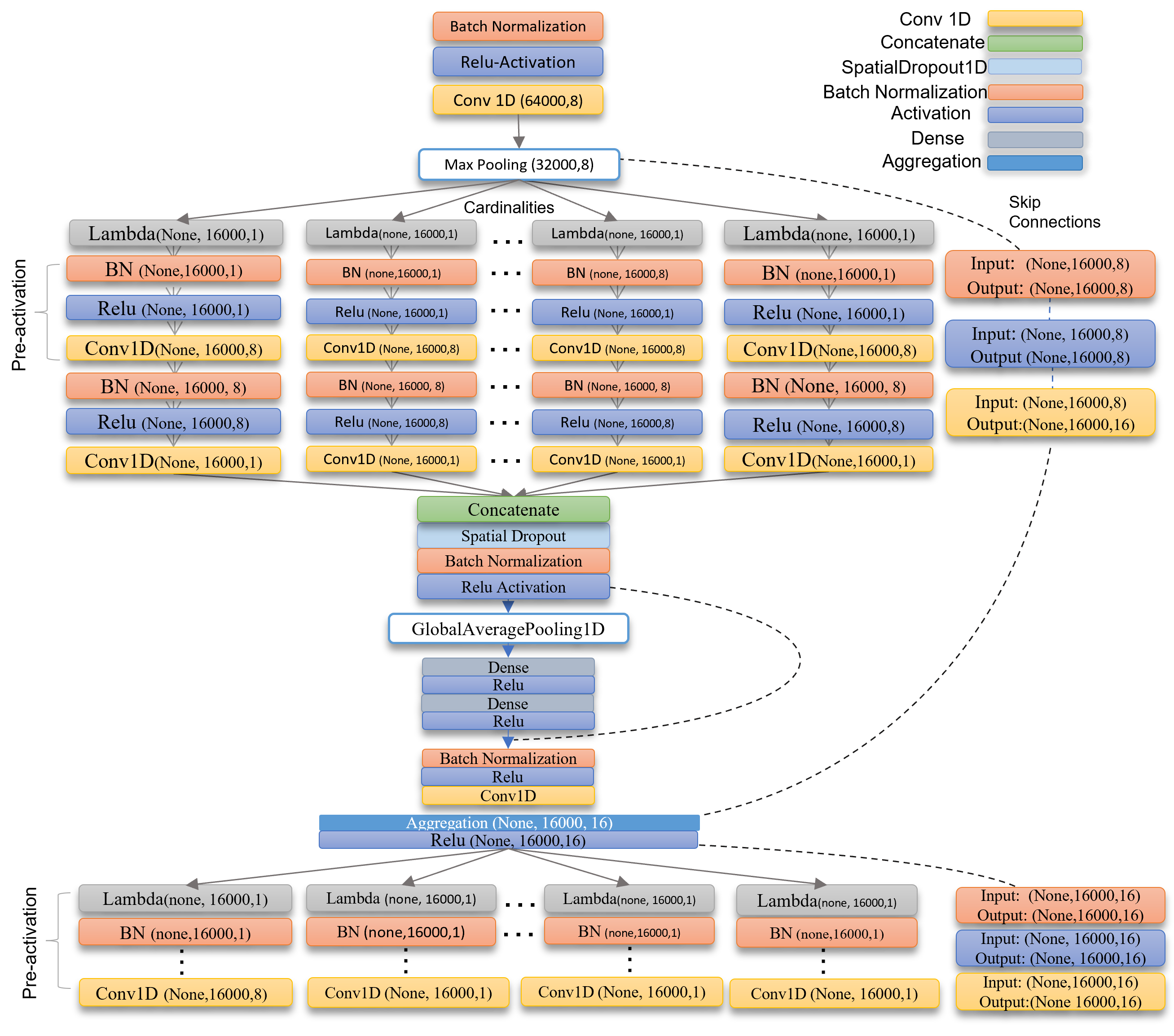}
        \caption{Intra-architecture of SE-PSA Blocks with 4 cardinalities and pre-activation convolutions. The similar Intra-architecture repeated 5 times for each block of the proposed PSA network.} \vspace{-10pt}
        \label{fig:Intra-architecture}
\end{figure*}
 
\begin{equation}
    S_R = \sum^{D}_{i=1}\omega_in_i
\end{equation}
where $n_i = \{n_1, n_2, \ldots,n_i$\} is the $D$ channel input vector, $\omega_in_i$ is the filter weight for the $i_{th}$ channel, and $S_R$ shows the inner product of the neural path layers. Analogous to a $S_R$, $\tau_i(n_i)$ projects $E^{st}_c$ embedding by applying the transformation to the convolved representation as shown below.
\begin{equation}
    E^{st}_c = \sum^{C}_{i=1} \tau_i(n_i)
\end{equation}
\begin{equation}
    F^{f}_r =  E^{st}_c + \sum^{C}_{i=1} \tau_i(n_i)
\end{equation}
where $F^{f}_r$ refers to the aggregated fine-grained embeddings extracted to classify the speech representations and $\tau_i(n_i)$ can be an arbitrary function. The aggregated representations from each SE-PSA block will be processed further by involving global max pooling and flattening. Along with these, we employed fully connected layers incorporating dropout to classify the speech sample as genuine or spoofed (Figure \ref{fig:spatialdropout}). For the classification, we used the sigmoid activation function with a threshold of 0.5, determining the final prediction as follows:

\begin{equation}
   \mathbb{S}_{cr}= \frac{1}{1+e^{-x}}
\end{equation}
\begin{equation}
    \mathbb{P}_{pred}=\mathbb{S}_{cr} > 0.5
\end{equation}
where $\mathbb S_{cr}$ denotes the score of the utterance and $\mathbb{P}_{pred}$ refers to the prediction of the model as spoof and bonafide speech. Further, to address network overfitting, spatial dropout is applied before the aggregation node. Spatial dropout involves selecting neurons with more significant embeddings and dropping those with less representative features, as illustrated in Figure~\ref{fig:spatialdropout}. Notably, PSA-Net incorporates squeeze-and-excitation (SE) blocks within each SE-PSA block as shown in Figure~\ref{fig:Squeeze&Excitation}. These blocks effectively capture transformed feature maps and enhance back-propagation, aiding in distinguishing subtle differences between replay-based and synthetic-based spoofing attacks. Thus, by combining multi-cardinality, dropout, SE blocks, and the split-transform-merge strategy, PSA-Net effectively differentiates between genuine and spoofed speech while reducing the impact of irrelevant parameters.


\subsubsection{Addressing Vanishing Gradients:}
While PSA-Net benefits from its reduced depth and wider architecture compared to deeper networks (like ResNet-101 or ResNet-51), achieving effective learning for different audio artifacts can still be affected by the vanishing gradient problem. This occurs when the gradient used to update network weights becomes increasingly smaller, making it difficult for multiple layers to learn complex patterns. Thus, instead of relying solely on parallel information flow, PSA-Net strategically incorporates skip connections. These connections directly bypass certain layers, adding the output of earlier layers to later ones (as shown in the equations below). Skip connections essentially inject the preserved gradient information directly into deeper layers, which enable PSA-Net to learn effectively even within multiple cardinalities. The inclusion of skip connections further enhances the ability of PSA-Net to discern authentic from spoofed speech. In particular, in a network without skip connections, the gradient is computed as follows:
\begin{equation}
    y = \frac{\partial J}{\partial x}
\end{equation}
where $y$ obtained using the chain rule for the full operation. In particular, the full operations can be performed as follows:
    \begin{equation}
 \frac{\partial J}{\partial x_0} =  \frac{\partial J}{\partial x_2} \frac{\partial x_2}{\partial z_2} \frac{\partial z_2}{\partial x_1} \frac{\partial x_1}{\partial z_1}\frac{\partial z_1}{\partial x_0}
\end{equation}
This denotes the chain of multiplications which renders neural networks prone to disappearing and exploding gradients. If we substitute F(x) for the intermediate computations, the gradient calculation becomes: 
 \begin{equation}
 \frac{\partial J}{\partial x} =  \frac{\partial J}{\partial F(x)} \frac{\partial F(x)}{\partial x} 
\end{equation}
Next, we add the new function $H(x) = F(x) + x$ for the skip connection. In particular, we must now differentiate $F(x)$ through $H(x)$ to get the gradient of the cost function in a network with skip connections as shown below: 
 \begin{equation}
 \frac{\partial J}{\partial x} =  \frac{\partial J}{\partial H(x)} \frac{\partial F(x)}{\partial x} 
\end{equation}
where the derivative of $x$ with respect to $ H(x)$ is equal to 1. Thus, substituting $F(x) + x$ for $H(x)$ yields the expression: 
 \begin{equation}
\frac{\partial J}{\partial x} = \frac{\partial J}{\partial H(x)} ({\frac{\partial F(x)}{\partial x}} + 1) =  \frac{\partial J}{\partial H(x)} \frac{\partial F(x)}{\partial x} + \frac{\partial J}{\partial H(x)} \end{equation}
In this scenario, the gradient of $F(x)$ becomes extremely small as a result of multiple matrix multiplications during back-propagation through all the layers of $x$. However, we still retain the direct gradient of the cost function concerning $H(x)$.
This approach allows the network to reduce certain gradient computations during back-propagation, preventing gradient vanishing or exploding.
In the next section, we outline the experimental setup for conducting experiments and comparative analyses of the proposed system.
 
\section{Experimental Setup}\label{Experiment}
This section describes the experimental configuration used to obtain the reported results. We evaluated the proposed PSA-Net against replay-based, voice cloning, and chained replay-based spoofing attacks. In addition, we describe the datasets, evaluation metrics, and hyper-parameter settings used for the experiments, testing, and analysis of the reported results.
\subsection{Dataset}
The ASVspoof challenges have been pivotal in advancing the development of VSDs by providing standardized datasets and benchmarks since 2015. Notably, the ASVspoof2019~\cite{todisco2019asvspoof} database has emerged as the standard benchmark for developing and evaluating VSDs. Consequently, in this paper, we used the widely adopted ASVspoof2019~\cite{todisco2019asvspoof} dataset by training the PSA-Net on the training subset and validating it on the development subset. The evaluation subset of ASVspoof2019 dataset was used for in-domain performance evaluation of PSA-Net against unseen spoofing attacks.

Additionally, in order to test the generalization ability of PSA-Net beyond ASVspoof2019, we tested the performance of PSA-Net against several recent databases like ASVspoof2021 (LA and PA)~\cite{yamagishi2021asvspoof}, PartialSpoof \cite{zhang2021initial} (Utterance-based LA only), and VSDC (Chained replay-based PA). The ASVspoof2021~\cite{yamagishi2021asvspoof} dataset includes spoofing audio generated by recent algorithms, representing both logical and physical access attacks. While the PartailSpoof dataset contains utterance-based logical access spoofing attacks based on frame-level spoofing injections (for further details, see \cite{zhang2021initial}). In contrast to LA and PA attacks, the VSDC~\cite{baumann2021voice} dataset is specifically composed of audio samples obtained from IoT devices and includes novel, chained replay-based forgeries. By including the VSDC dataset, we ensure a thorough evaluation of PSA-Net's effectiveness in the context of IoT-enabled devices. Further details on the configurations, samples, and speakers, etc., of these datasets are provided in Tables \ref{tab:Asvspoof2019}, \ref{tab:VSDC}, and in studies \cite{zhang2021initial, yamagishi2021asvspoof,baumann2021voice,todisco2019asvspoof}.

\begin{table}[t]
\caption{ Statistical details of ASVspoof Challenge 2019 database.}\vspace{-5pt}
\label{tab:Asvspoof2019}
\begin{tabular}{ccccccc}
\toprule
  & \multicolumn{2}{c}{Speaker} & \multicolumn{2}{c}{LA Attacks}  & \multicolumn{2}{c}{PA Attacks}  \\  
 Subset & Male  & Female & Genuine & Spoofed & Genuine & Spoofed \\
\midrule
 Training & 8 & 12 & 2580 & 22800 & 5400 & 48600\\
 Development & 8 & 12 & 2548 & 22296 & 5400 & 24300\\ \hline
 Evaluation & - & - & \multicolumn{2}{c}{71747} & \multicolumn{2}{c}{137457}  \\
\bottomrule
\end{tabular}
 \end{table}\vspace{-5pt}

\vspace{-5pt}\begin{table}[b]
\caption{Summary of VSDC Database.}
\label{tab:VSDC}
\centering
\begin{tabular}{cccc}
\toprule
  Audio samples & Speech Samples  & Environment & Sample Rate\\
\midrule
 Bonafide & 4000 & \multirow{4}{2.5cm}{Recording Chamber, Kitchen Table, Living Room, Office Desk} & \multirow{4}{1cm}{96K}\\
 Replay & 4000 &  & \\
 Clonded Replay & 4000 &  &  \\
 Total & 12000 &  & \\
\bottomrule
\end{tabular}
 \end{table}
 
\subsection{Data Preprocessing}
\label{pp}
Raw audio signals, characterized by fluctuating high- and low-frequency values require preparation before being fed into the PSA-Net. This preprocessing stage involves two steps: frame length adjustment and normalization.
\subsubsection{Frame Length Adjustment:}\vspace{-5pt}
For datasets like ASVspoof2019, where audio files have varying lengths and frame counts, we standardize the input voice sample length $L$ to four seconds. This is achieved by either padding or trimming the voice sample. For shorter samples, we retain the first four seconds (equivalent to $1 \times 64000$ values) and neglect the remaining length of the audio. Conversely, for longer samples, we truncate any excess audio beyond the four-second window.
\subsubsection{Z-score Normalization:}\vspace{-5pt}
Subsequently, Z-score normalization is applied to the audio waveform. This process constrains the sample values to the range [$−1,1$] based on the mean and standard deviation of the signal. The mathematical computation we applied for this transformation is as follows:

\begin{equation}
    x = \frac{\sum_{i=1}^{j} (x - \mu)}{\sigma} 
\end{equation}
\begin{equation}
    \sigma = \sqrt{E[X^2]-(E[X])^2} \\
\end{equation}
where $x$ is the number of speech samples, $\mu$ and $\sigma$ are the mean and standard deviation of the signal. After preprocessing, we apply five different types of augmentation to address the data imbalance issues. The details of data augmentation are explained in the next subsection below.
\begin{figure}[!b]
        \centering
        \includegraphics[width=15cm]{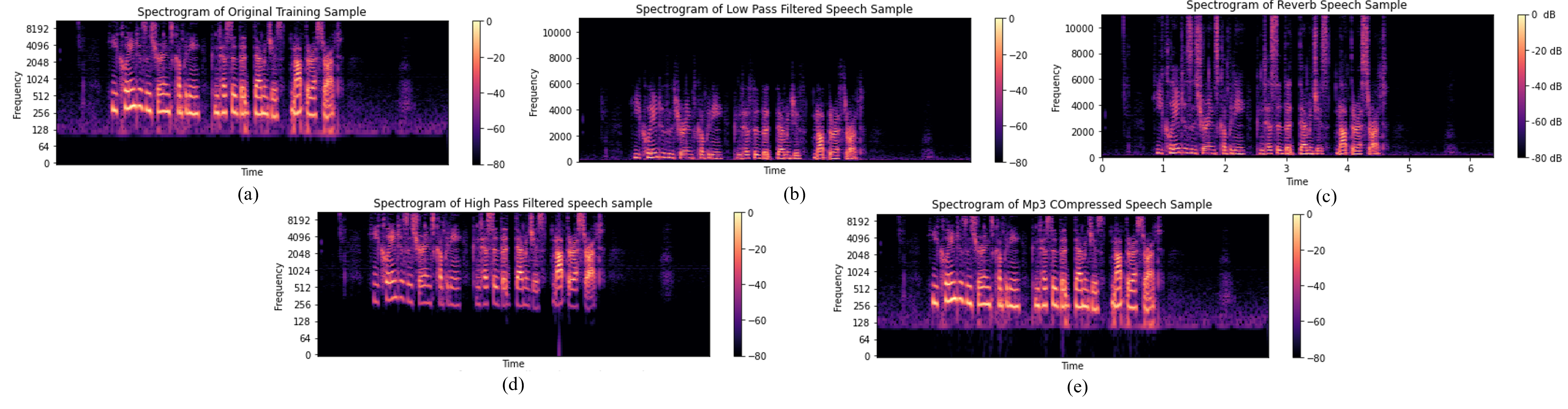}\vspace{-5pt}
        \caption{The spectrogramatic representation of different augmentation applied to remove class imbalance and enhanced the learning of the model from broader frequency spectrum: (a) Original signal, (b) High pass Filtering, (c) Mp3 compression, (d) Low pass filtering, (e) Reverberation }\vspace{-5pt}
        \label{fig:spectrogram}
\end{figure}

\subsection{Data Augmentation}
\label{DA}
Data augmentation (DA) serves as a crucial technique in both image and speech recognition, aiding in the amplification of training data, reducing overfitting, and enhancing performance, particularly in scenarios with imbalanced classes~\cite{loshchilov2017decoupled}. To address these challenges within this paper, we implemented five effective augmentation techniques: MP3 compression, high-pass filtering, low-pass filtering, silence trimming, and reverberation.

MP3 compression has shown effectiveness in spoofing detection~\cite{tak2022automatic}, and the other augmentations were selected based on their positive impact on model performance. High- and low-pass filtering, specifically, aid in extracting sub-band information, crucial for discerning fine-grained features in contemporary spoofing attacks. Additionally, we included reverberation, aligning with the configuration settings used during dataset preparation for evaluation. The spectral analyses of all five augmented techniques over speech samples are depicted in Figure~\ref{fig:spectrogram}. The spectra of the speech samples exhibit that the model was trained with diverse frequency sub-bands, which assists in learning the diverse artifacts of the speech samples.

\subsection{Evaluation Metrics}
\label{Eval_Met}
From the dataset details in Table~\ref{tab:Asvspoof2019}, it is clear that the ratio of real to spoofed trials is highly skewed. To address this, we employ alternative performance metrics, namely the Equal Error Rate (EER) and the Tandem Detection Cost Function (t-DCF), which are standard in ASVspoof challenges. In contrast to EER, t-DCF measures the performance evaluation of spoofing countermeasures (CMs) on the reliability of an ASV system. Wang et al.~\cite{wang2021comparative} demonstrate that the effectiveness of spoofing detection systems can vary significantly when random seeds are used. Similarly, after being trained with various random seeds, the EER of the baseline system in~\cite{tak2021end} fluctuates between $1.19\%$ and $2.06\%$. Furthermore, to evaluate the compatibility of the proposed PSA-Net with IoT enabled devices, we use FLOPs matrix along with inferencing time and other parameters to provide insights from IoT aspect. 

\subsection{Configurations and Hyper-parameters}
\label{ES&HP}
For all experiments, we utilize the Keras training platform in Python. The PSA-Net employs Adam optimizer with an initial learning rate of $1e^{-4}$ and a weight decay of $0.001$. The filter and kernel values for convolution layers are set to $64,128,256$ and $196,144,100$ for $c1,c2,c3$ and $k1,k2,k3$, respectively. We adopt the cosine annealing warm restarts method ~\cite{loshchilov2017decoupled} to adjust the learning rate, with linear growth for the first $1000$ warm-up steps, followed by a decrease according to the inverse square root of the step number. The PSA-Net model is trained for $50$ epochs using the cross-entropy loss function. Further, KAIMING initialization~\cite{he2015delving} is employed for all convolution layers, and batch normalization layers are configured with weights at $1$ and biases at $0$. The final model for evaluation is chosen based on the lowest loss observed on the development set. The training and testing of the proposed PSA-Net is done on the Matilda High Performance Cluster at Oakland University, Michigan, USA. The HPC's GPU nodes, equipped with four NVIDIA Tesla V$100$ $16$GB GPUs, $192$ GB of RAM, and $48$, $2.10$GHz CPU cores, were utilized for these tasks.

\section{Results and Discussions}\label{results}
This section presents the experimental and comparative analysis of the proposed PSA-Net for voice-spoofing attack detection. We aimed to evaluate the effectiveness of PSA-Net against various spoofing attacks within four different datasets, including ASVspoof2019~\cite{todisco2019asvspoof}, ASVspoof2021~\cite{yamagishi2021asvspoof}, PartialSpoof~\cite{zhang2021initial}, and VSDC~\cite{baumann2021voice}. We optimized the hyperparameters of the PSA-Net model (details in Section~\ref{ES&HP}) and present the results achieved with the best configuration below. Furthermore, we compare the performance of PSA-Net with 11 existing standalone and 7 unified voice-spoofing detection methods using standard evaluation metrics (details in Section~\ref{Eval_Met}). Our results demonstrate that PSA-Net achieves significantly better performance on all datasets compared to existing methods, highlighting its effectiveness and potential for real-world IoT enabled applications. In addition to the experimental analysis, we also present a detailed ablation study and discuss the trade-off between differently configured PSA-Net with varying cardinalities and widths of the models. 

\subsection{Trade-off between Cardinalities and Bottleneck Depth}
In networks employing split-aggregate methodologies with multiple pathways, two key parameters are cardinality $(C)$ and bottleneck depth $(d)$. Unlike Inception networks with unique cardinal paths, the proposed PSA-Net employs diverse cardinalities but identical configurations for each pathway. Inspired from high cardinalities with proven optimal in aggregation based networks for large image datasets like ImageNet and CIFAR~\cite{xie2017aggregated}, we explore the trade-off between cardinality $C$ and bottleneck width $d$. 

\begin{table}[b]
\caption{Performance analysis of the proposed SE-PSA network with different cardinalities (\textit{C}) and model width (\textit{d}) to show the trade off in model effectiveness with ASVspoof2019 balanced speech samples.}
\label{tab:tradeoffcard}
\centering
\begin{tabular}{cccc}
\toprule
  Cardinality (\textit{C}) & Model-Width  (\textit{d})  & AUC-LA & AUC-PA \\
\midrule
 1             & 64 &  0.78 & 0.79 \\
 2              & 40 & 0.61 & 0.59  \\
 4              & 32   & 0.92 & 0.93 \\
  \textbf{4}     & \textbf{64} &\textbf{ 0.93 }&  \textbf{0.97}\\
 8              & 32  & 0.92 &  0.93\\
 8              & 64  & 0.73 & 0.67\\
 16              & 4   & 0.87 & 0.89\\
\bottomrule    
\end{tabular}
 \end{table}

 
In this experiment we uses a balanced set of genuine and spoofed speech samples from both LA and PA subsets of the ASVspoof2019 dataset. As results demonstrated in Table \ref{tab:tradeoffcard}, the $4C\times64d$ configuration outperforms other variants with Area Under the Curve (AUC) scores of $0.93\%$ and $0.97\%$ for LA and PA subsets, respectively. Notably, the AUC fluctuates as $C$ increases from $1$ to $32$. Interestingly, setting $C$ to $1$ reduces the proposed network to a standard ResNet. For the initial values of $C$ and $d$, we chose the values proven to be effective against large-scale datasets~\cite{xie2017aggregated}.\\
Further examination indicates that the $8\textit{C}\times32\textit{d}$ and $4\textit{C}\times24\textit{d}$ structures yield comparably second-best results, although their performance degrades when detecting replay spoofing, particularly in AUC scores. Further, the analysis demonstrates that as bottleneck width decreases and cardinality increases, the AUC starts to saturate. Thus, while prior studies advocated higher cardinalities and widths \cite{xie2017aggregated}, our findings suggest that, for the ASVspoof2019 dataset, optimal results are achieved within cardinality ranges from $4$ to $8$ and width ranges between $32$ and $64$. Consequently, the most effective performance exhibits by PSA-Net obtained with cardinality of $4$ and a model width of $64$.

\begin{table}[b]
\caption{Performance analysis of the proposed SE-PSA network against different spoofing attacks from diverse datasets}\vspace{-5pt}
\label{tab:performance-proposed}
\centering
\begin{tabular}{*{6}{c}}
\toprule
  \multirow{2}{2cm}{Datasets} & \multicolumn{2}{c}{Cloning} & \multicolumn{2}{c}{Replay} & Chained Replay  \\ 
                         & EER & t-DCF & EER &t-DCF & EER  \\
\midrule
 ASVspoof 2019 \cite{todisco2019asvspoof} & 3.04 & 0.087 & 1.26 & 0.038 & --  \\
 ASVspoof 2021 \cite{yamagishi2021asvspoof} & 3.78 & 0.135 & 3.40 & 0.129 & --  \\
 Partial spoof \cite{zhang2021initial} & 6.30 & 0.227 & -- & -- & --  \\
 VSDC \cite{baumann2021voice} & -- & -- & 0.32 & -- & 0.87  \\
\bottomrule   
\end{tabular}
 \end{table}

\subsection{Performance Analysis of the Proposed PSA-Net Against Different Voice-Spoofing Attacks}

\subsubsection{Performance Analysis Against Different Voice-spoofing Datasets:}
In this experiment, we evaluate the performance of the proposed PSA-Net solution on different datasets: the ASVspoof2019~\cite{todisco2019asvspoof}, VSDC~\cite{baumann2021voice}, Partialspoof~\cite{zhang2021initial}, and ASVspoof2021~\cite{yamagishi2021asvspoof}. We use the training subsets of the datasets (ASVspoof2019, PartialSpoof, and VSDC) for training, the development subsets are used for validation, and the evaluation subsets are used to test the effectiveness of the system. The ASVspoof2021 data is only used to test the performance of the PSA-Net against recent spoofing algorithms, as there are no training files available for this dataset. The results of the evaluation are presented in Table \ref{tab:performance-proposed}, where the proposed system achieves impressive performance, with an EER of 3.04\% for LA and 1.26\% for PA and minimal t-DCFs of 0.087 and 0.038, respectively. In the case of the VSDC dataset, the proposed system obtains an EER of 0.32\% for the first-order replays and an ideal EER of 0.87\% when the speech sample contains the artifacts of multi-order replays. While for the PartialSpoof datasets, the PSA-Net obtained an EER of 6.30\%, which demonstrates the effectiveness of our PSA-Net against frame-level-based spoofing. The results show that the proposed system performs better when tested against replay and chained replay spoofing attacks. This demonstrates the system's effectiveness against device artifacts in playback voice samples. Although the proposed system performs marginally better in PA attacks, it still performs better in cloning attack detection in comparison to existing solutions.

\subsubsection{Performance Analysis Against Unified Training of SOTA Spoofing Attacks:}
In this sub-experiment, we evaluate the effectiveness of the PSA-Net against unseen attacks without prior knowledge of the specific spoofing type, simulating real-world scenarios. While existing literature lacks evaluations on unified spoofing classes, where the performance of their models are separate tested on clone and replay attacks sequentially. To address this gap, we created an integrated spoofing category by combining both LA and PA voice samples from the ASVspoof2019 and VSDC datasets. The proposed model is then trained to simultaneously distinguish between bonafide, clone, and replay samples within this unified category. During testing, the model encounters both LA and PA attacks simultaneously, achieving an EER of 5.35\% and a t-DCF of 0.237. While slightly higher results were obtained with known attack types (3.04\% and 0.087 EER/t-DCF for LA in Table \ref{tab:performance-proposed}), this demonstrates the applicability of PSA-Net to real-world spoofing challenges. More importantly, these results highlight the limitations of previous research, where separate training and testing fail to accurately assess system performance against real-time diverse attacking scenarios. Specifically, restricting training and evaluation to either LA- or PA-based attacks significantly degrades model performance when facing multiple spoofing attacks in real-time practical environments. Therefore, This unique training approach not only showcases the real-world applicability of PSA-Net but also exposes the limitations of existing separate training approaches. Further, this experimental approach enabled the development of more robust and generalize VSD systems.



\begin{table}[b]
\caption{Performance comparison of the proposed PSA-Net with SOTA comparative methods against ASVspoof2019 synthesized and converted speech samples, where * refers to training with augmented samples.}\vspace{-5pt}
\label{tab:LAresults}
\centering
\begin{tabular}{cccc}
\toprule
  Method & Input   & EER & min T-DCF \\
\midrule
 Baseline-Asvspoof2019 \cite{todisco2019asvspoof} & CQCC+GMM & 9.57 & 0.237  \\
 Baseline-Asvspoof2019 \cite{todisco2019asvspoof}& LFCC+GMM  & 8.09 & 0.212 \\
 Chettri et al. \cite{chettri2019ensemble} & Spatial features & 7.66 & 0.179\\
 Monterio et al. \cite{monteiro2020generalized} & Spectrogram  & 6.38 & 0.142\\
 Gomez-Alanis et al.\cite{gomez2019light} & Spectrogram  & 6.28 &  -\\
 Aravind et al.    \cite{aravind2020audio} & Mel-spectrogram & 5.32 & 0.151\\
 Lavrentyeva et al.  \cite{lavrentyeva2019stc}   & CQT & 4.53 & 0.103\\
 ResNet + OC-SVM  \cite{zhang2021one}  & -- & 4.44 & 0.115\\
 Wu et al  \cite{wu2020light}  & deep features & 4.07 & 0.102\\
 Tak et al.  \cite{tak2020spoofing}   & LFCC & 3.50 & 0.090\\
 Chen et al.  \cite{chen2020generalization}  & Filter banks & 3.49 & 0.092\\
 One class Learning \cite{zhang2021one}& LFCC-60D &  2.19 & 0.059 \\
\textbf{ PSA-18 layers }   & Raw audio & 4.06 & 0.099\\
\textbf{PSA-18 layers*}&\textbf{Raw audio} & \textbf{3.04 }& \textbf{0.087} \\
\bottomrule   
\end{tabular}
 \end{table}
\subsection{Comparative Performance Analysis of the Proposed PSA-NET Against Voice Cloning Attacks}
In this experiment, we examine the performance of the proposed PSA-Net against voice cloning (synthetic and converted) voice-spoofing attacks. The model is trained using the training subset of the ASVspoof2019-LA dataset, along with five types of augmented samples as described in Section~\ref{Experiment}. To evaluate the performance of our presented solution, the PSA-Net is compared with twelve state-of-the-art methods, and the results are presented in Table \ref{tab:LAresults}. The results demonstrate that, when trained using augmented samples, the proposed system obtains an EER of 3.04\% and a t-DCF of 0.087, whereas without augmented samples, the model achieves 4.06\% and 0.099, respectively. These results indicate that the proposed system outperformed eleven out of the twelve SOTA comparative countermeasures, with the lowest EER and t-DCF. More specifically, the proposed system performed second best on the ASVspoof2019-LA dataset, both with and without augmented samples. Besides having a slightly lower EER and minimal t-DCF than the presented system in \cite{zhang2021one}, the system is particularly optimized to identify only LA-based attacks and has never been evaluated against replay attacks. In contrast, the proposed system obtained a lower EER and minimal t-DCF even when trained without any type of augmentation. This indicates the proposed system's superiority as a robust countermeasure to voice cloning and conversion attacks. 
\begin{table}[t]
\caption{\small Comparison of proposed method with existing methods on ASVSpoof2021 dataset (Lower is better).}\vspace{-5pt}
\label{tab:asvspoof2021}
\centering
\begin{tabular}{cccc}
\toprule
\multirow{2}{1cm}{Study} & \multirow{2}{1cm}{Method} & \multicolumn{2}{c}{ASV-21} \\
 &  &   LA & PA \\
   \midrule
         \cite{tak2022automatic}  & wav2vec 2.0 & 1.19 & 4.38   \\
          \cite{tomilov21_asvspoof}  & LCNN+ResNet+RawNet & 1.32 & 15.64  \\ 
        \cite{das2021known}  & GMM+LCNN (Ensemble) & 3.62 & 18.30  \\
           \cite{chen2021ur} & ECAPA-TDNN (Ensemble) & 5.46 & 20.33   \\ 
        \cite{chen2021pindrop} & ResNet (Ensemble) & 3.21 & 16.05  \\
           \cite{wang2021investigating}  & W2V2 (fixed)+LCNN+BLSTM & 10.97& 7.14   \\      
           \cite{wang2021investigating}  & W2V2 (finetuned)+LCNN+BLSTM & 7.18 & 5.44   \\
           \cite{martin2022vicomtech} & FIR-WB & 3.54 & 4.98   \\ 
            \textbf{PSA-Net}  & Raw audio & \textbf{3.78} & \textbf{3.40}  \\    
\bottomrule 
\end{tabular}
\end{table}
\begin{table}[b]
\caption{\small Comparison of proposed method with existing methods on Partial spoof dataset (Lower is better).}\vspace{-5pt}
\label{tab:partialspoof}
\centering
\begin{tabular}{ccc}
\toprule
Study & Method & EER \\
 \midrule
         \cite{zhu2023local}  & LCNN & 6.19 \\
          \cite{zhu2023local}  & SELCNN & 6.33   \\ 
        \cite{zhu2023local}  & H-MIL (Ensemble) & 5.96  \\
           \cite{zhu2023local} & LS-H-MIL & 5.89   \\ 
        \cite{zhang2021multi} & LCNN + LSTM & 8.61   \\
           \cite{zhang2021multi}  & SELCNN(2)+LSTM & 7.69 \\      
            \textbf{PSA-Net} & Raw audio & \textbf{6.30}  \\    
 \bottomrule 
\end{tabular}
\end{table}
Furthermore, to test the generalization of PSA-Net, we evaluated its performance against synthetic and converted voices developed with unseen algorithms and approaches. Specifically, we compared PSA-Net's performance on the ASVspoof2021~\cite{yamagishi2021asvspoof} and PartialSpoof~\cite{zhang2021initial} datasets. The results, presented in Tables~\ref{tab:asvspoof2021} and \ref{tab:partialspoof}, show that PSA-Net achieves an EER of 3.78\% for LA attacks and 3.40\% for replay attacks on the ASVspoof2021 dataset, even though the testing samples are unseen by the model. Similarly, on the PartialSpoof dataset, PSA-Net outperforms four out of six comparative methods evaluated in \cite{zhu2023local} and \cite{zhang2021multi}, as shown in Table~\ref{tab:partialspoof}. While PSA-Net was not the best-performing method in these generalization tests, it effectively detected other types of voice-spoofing attacks, which is missing in the best-performing solutions. Thus, based on the overall results of logical voice-spoofing and generalization tests, PSA-Net has proven to be an optimal solution. 

\subsection{Comparative Performance Analysis of the Proposed PSA-Net Against Replay Spoofing Attacks}
In this experiment, we test the effectiveness of the proposed PSA-Net solution against replay spoofing attacks. The proposed system is trained using the training subset of the ASVspoof2019-PA dataset, validated using the development subset, and evaluated against the evaluation subset of the ASVspoof2019-PA dataset. The results of this experiment are presented in Table \ref{tab:PAresults}. The results indicate that the proposed system effectively discriminates between spoofed and bona fide artifacts in replayed voice samples. When trained using augmented samples, the proposed system achieves an optimal EER of 1.26\% and a minimal t-DCF of 0.038. In comparative analysis to the other existing solutions, the proposed PSA-Net attains an EER of 2.13\% and a t-DCF of 0.064 without using the augmentation samples during training. Further, the results show that the proposed system achieved the second-lowest EER after ASSERT~\cite{lai2019assert}. Although the EER of the ASSERT~\cite{lai2019assert} solution is slightly better in PA spoofing detection, the proposed model outperformed ASSERT~\cite{lai2019assert} in LA spoofing attack detection. Moreover, the ASSERT~\cite{lai2019assert} model is based on handmade features and a 50-layer SENet architecture, whereas the proposed model has an 18-layer architecture and can extract the required deep features directly from raw audio. Therefore, at the end of this experiment, we can conclude that, except for the ASSERT, the proposed PSA-Net model outperformed the other eight comparative models in replay speech detection and has proven to be effective in replay attack detection. 
\begin{table}[t]
\caption{Performance comparison of the proposed PSA network with SOTA comparative methods against Replay Attacks; * refers to training with augmented samples.}
\label{tab:PAresults}\vspace{-5pt}
\centering
\begin{tabular}{cccc}
\toprule
  Method & Input   & EER & min T-DCF \\
\midrule
 Baseline-Asvspoof2019 \cite{todisco2019asvspoof} & CQCC+GMM & 11.04 & 0.2454  \\
 Baseline-Asvspoof2019 \cite{todisco2019asvspoof}& LFCC+GMM  & 13.54 & 0.3017 \\
 ASSERT \cite{lai2019assert} & log-spec-SENet & 1.29 & 0.036\\
 ASSERT \cite{lai2019assert} & SENet50-Dialated ResNet  & 0.59 & 0.016\\
 STC  \cite{lavrentyeva2019stc}   & LFCC-CMVN-LCNN & 4.6 & 0.105\\
 STC  \cite{lavrentyeva2019stc}   & FFT-LCNN & 2.06 & 0.56\\
 BUT-Omilia  \cite{zeinali2019detecting}  & logSpec-VGG-SincNet 1
-SincNet 2 & 1.51 & 0.0372\\
 BUT-Omilia  \cite{zeinali2019detecting}  & SincNet with standard
dropout & 2.11 & 0.052\\
 BUT-Omilia  \cite{zeinali2019detecting}  & VGG 1-VGG 2 & 1.49 & 0.04\\
 BUT-Omilia  \cite{zeinali2019detecting}  & SincNet with high dropout & 2.31 & 0.059\\
\textbf{PSA-18 layers  }   & Raw audio & 2.13 & 0.064\\
\textbf{PSA-18 layers*}&\textbf{Raw audio} & \textbf{1.26}& \textbf{0.038} \\
\bottomrule  
\end{tabular}
 \end{table}

 
\begin{table}[b]
\caption{Experimental performance of the proposed system against SOTA unified voice-spoofing countermeasures on ASVspoof2019 corpus.}\vspace{-5pt}
\label{tab:unifiedresults}
\centering
\begin{tabular}{ccccc} 
 \toprule
 \multirow{2}{1cm}{Paper}  & \multicolumn{2}{c}{Logical Access} & \multicolumn{2}{c}{Physical Access} \\ 
                         & EER & min T-DCF & EER & min T-DCF  \\
 \midrule
 Baseline \cite{todisco2019asvspoof}& 11.96 & 0.212 & 13.54  & 0.3017    \\
 Baseline \cite{todisco2019asvspoof} & 9.87 & 0.236 & 11.04  & 0.2454   \\
 ASSERT \cite{lai2019assert}  & 11.75 & 0.216 & 1.29  & 0.036  \\
 ASSERT \cite{lai2019assert}  & 6.70 & 0.155 & 0.59  & 0.016  \\
 STC  \cite{lavrentyeva2019stc}   & 7.86  & 0.183 & 4.6  & 0.105  \\
 STC  \cite{lavrentyeva2019stc}  & 4.53 & 0.103 & 2.06 & 0.56 \\
 BUT-Omilia  \cite{zeinali2019detecting} & 8.01 &0.208& 1.51& 0.0372  \\
 BUT-Omilia  \cite{zeinali2019detecting}  & 8.01 &0.356 &2.11& 0.0527 \\
 BUT-Omilia  \cite{zeinali2019detecting}  & 10.52 &0.279& 1.49& 0.04 \\
 BUT-Omilia  \cite{zeinali2019detecting}  & 22.99& 0.381& 2.31 &0.0591 \\
MFMT \cite{li2019anti}   & 7.63 &0.213 &0.96 &0.0266 \\
 SASV  \cite{aljasem2021secure} & 5.22& 0.132 & 1.1 &0.0335 \\
 \textbf{PSA-18 layers}     & 4.06 & 0.099&  2.13 & 0.064  \\
\textbf{PSA-18 layers*}& \textbf{3.04 }& \textbf{0.087} & \textbf{1.26}& \textbf{0.038} \\
\bottomrule 
\end{tabular}
\end{table}
\subsection{Cumulative Performance Analysis of Proposed PSA-Net Against Comparative Unified Solutions}
In this experiment, we evaluated the performance of our proposed PSA-Net against SOTA unified solutions for detecting synthetic and replay-based attacks within a single system, and the results are presented in Table~\ref{tab:unifiedresults}. The results demonstrated that our method notably outperformed other six existing solutions in terms of EER and minimal t-DCF metrics.
\begin{figure}[t]
  \centering
  \begin{subfigure}[b]{0.45\textwidth}
    \includegraphics[width=7cm, height=5cm]{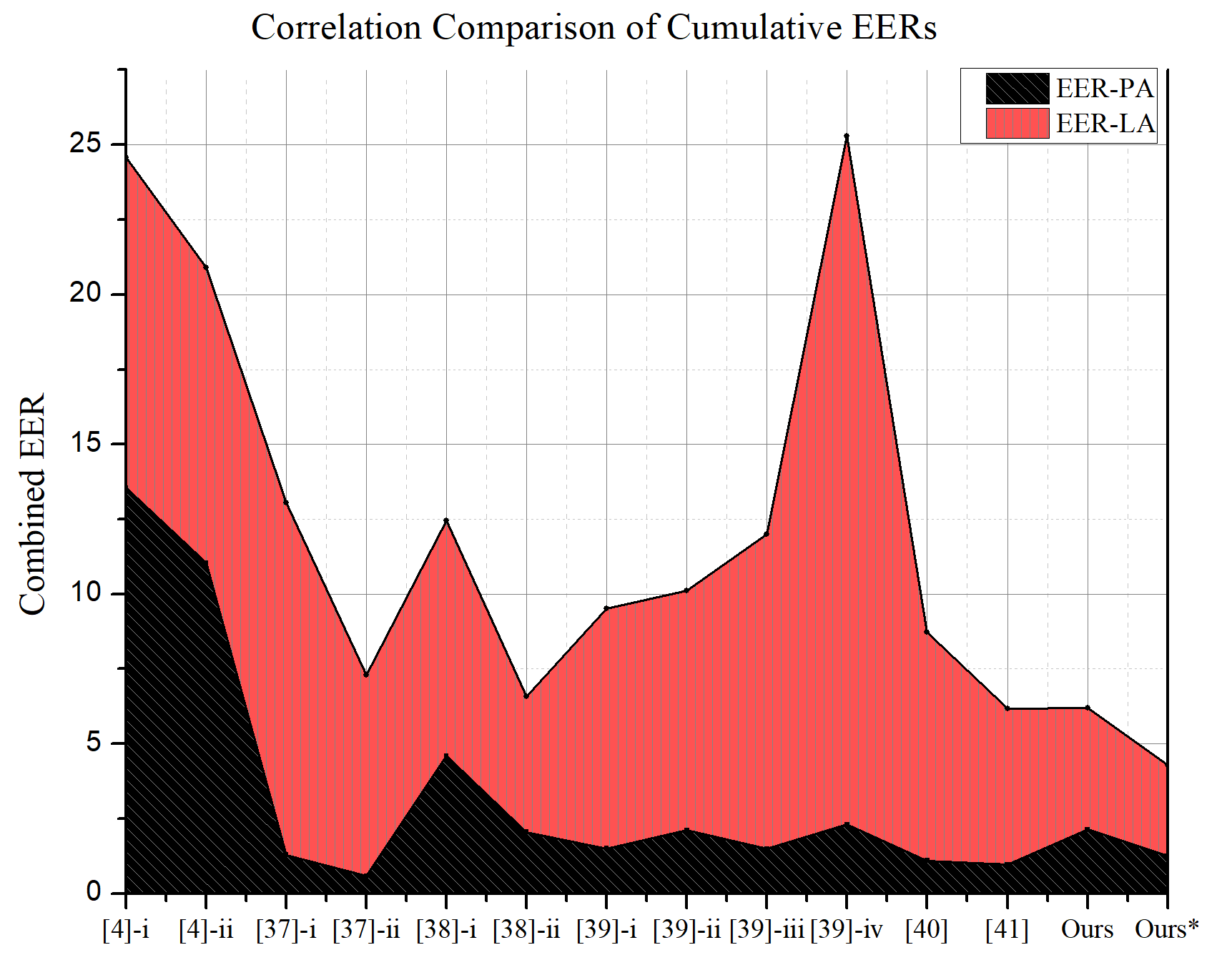}
    \caption{Cumulative EER comparison}
    \label{fig:commulativeEER-sub}
  \end{subfigure}
  \hfill
  \begin{subfigure}[b]{0.45\textwidth}
    \includegraphics[width=7cm, height= 5cm]{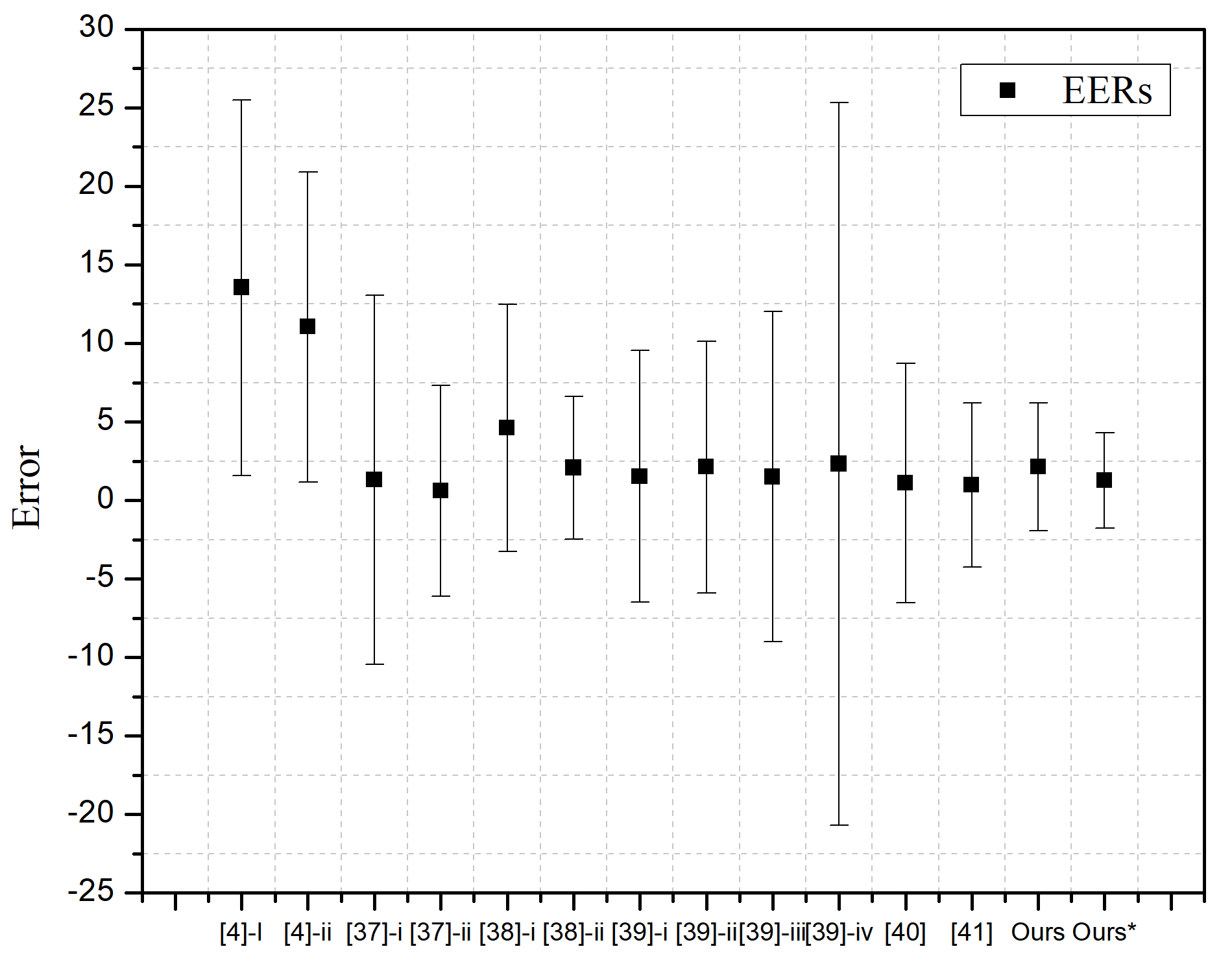}
    \caption{Error bar graph of EER}
    \label{fig:ERRofEER-sub}
  \end{subfigure}
  \caption{Comparison of existing deepfake detection methods. * denotes augmentation. [37]-i and ii denote SENET-ASSERTS and SENET-Resnet variation, [38]-i and ii denote FFT-CNN and LFCC-CNCC combination, and [39] i-iv denote logspec, SincNet, VGG, and SincNet with dropout, respectively.}
  \label{fig:combined}
\end{figure}

With data augmentation on the ASVspoof2019 dataset, PSA-Net achieved the lowest EER of 3.04\% and t-DCF of 0.087 for both synthetic and converted speech samples. Even without augmentation, PSA-Net maintained robust performance with EER and t-DCF values of 4.06\% and 0.099, respectively, especially against voice cloning attacks. Furthermore, PSA-Net outperformed other existing solutions for replay-based spoofing detection in the ASVspoof2019 dataset. With augmentation, it achieved an EER of 1.26\% and a t-DCF of 0.038, and without augmentation, the values were 2.13\% and 0.064, respectively. To further illustrate the overall performance, we compared the cumulative EER of PSA-Net with existing methods. This metric combines the EERs of different systems to evaluate their effectiveness against both logical and physical attacks. In this analysis, PSA-Net achieved an impressively lower cumulative EER of nearly 4.30\%, surpassing other unified methods, as shown in Figure~\ref{fig:combined}-(a). The figure clearly demonstrates that PSA-Net with augmentation (red and black bars) has the lowest EER overall. Additionally, the box plot in Figure~\ref{fig:combined}-(b) visualizes the error bars for each method. The error bars clearly show that PSA-Net outperforms other unified approaches against both LA and PA attacks. Overall, the experiment demonstrates that the proposed PSA-Net performs collectively better than all other unified solutions. Notably, unlike PSA-Net's end-to-end approach, other existing systems rely on computationally expensive handcrafted feature extraction. This further highlights the superiority of PSA-Net as a state-of-the-art unified countermeasure against LA and PA attacks.


\subsection{Performance Comparison of PSA-Net Against Traditional Handcrafted Features}
In this experiment, we tested the optimal input for our PSA-Net model for effective voice-spoofing detection. For this, we feed the traditional handcrafted features into the proposed SE-PSA network along with raw audio samples, and the results are reported in Table~\ref{tab:handcraft}. In prior studies~\cite{khan2023battling}, the handcrafted features have proven to be successful in voice spoof detection; however, their efficacy within the aggregated network has yet to be investigated. In order to demonstrate the efficiency of the PSA network against handcrafted features, we examined five commonly used handcrafted features for voice-spoofing detection: CQCC~\cite{todisco2016new}, LFCC~\cite{sahidullah2015comparison}, MFCC~\cite{zheng2001comparison}, GTCC~\cite{javed2022voice}, and LPCC~\cite{sahidullah2015comparison}. A $20$-number filter is used for all features except for CQCC~\cite{todisco2016new}, where a 96-octave filter is used. We use mean aggregation to transform the 2D feature specifications into 1D before feeding them into the network. For this experiment, we examine the performance of the PSA-Net with a balanced subset of the ASVspoof2019-LA dataset, in which we randomly select an equal number of real and spoof samples. The results in Table~\ref{tab:handcraft} indicate that the aggregated network incorporates the higher-order distinctions of the feature map fed into it. The local and global transformations of the feature map need a larger input to be performed optimally. As a result, the raw wave input with the appropriate number of frame lengths performs better than the other handcrafted features. The results show that the LFCC~\cite{sahidullah2015comparison} and GTCC~\cite{javed2022voice} features perform overall better, with an EER of $10.09\%$ and $7.66\%$, respectively; however, this EER is higher than raw wave audio. In contrast, the MFCC~\cite{zheng2001comparison} features show the highest EER, $26.38\%$, and a minimum t-DCF of $0.352$. Thus, we can conclude that the designed aggregated network performs better with raw waveforms compared to handcrafted features. \vspace{-5pt}

 \begin{table}
 \caption{Experimental performance comparison of the PSA-Net with raw audio and traditional handcrafted features.}
\label{tab:handcraft}
\centering
\begin{tabular}{c c c}
 \toprule
  Input   & EER & min t-DCF \\ 
  \midrule
 CQCC~\cite{todisco2016new} & 11.57 & 0.317  \\
 LFCC~\cite{sahidullah2015comparison}  & 10.09 & 0.292 \\
 GTCC~\cite{javed2022voice} & 7.66 & 0.179\\
 MFCC~\cite{zheng2001comparison}  & 26.38 & 0.352\\
 LPCC~\cite{sahidullah2015comparison}  & 16.38 & 0.242\\
 \textbf{Raw audio} & \textbf{2.65} & \textbf{0.079}\\
\bottomrule  
\end{tabular}
 \end{table}

\begin{table}[b]
\caption{Time complexity and compatibility analysis for IoT enabled devices}
\label{tab:time_complxity}
\centering
\begin{tabular}{ccccccc}
\toprule \\
Model (\textit{C} $\times$ \textit{d}) & Parameters (M) & Size(mb) & FLOPs(G) & \multicolumn{3}{c}{Inference time (sec)} \\ \midrule
 &             &          &         & GPU     & CPU       & Jetson-Nano          \\
PSA-1  (4 $\times$ 32)   & 25.30   & 24.10 & 2.21  &  0.90$\pm$0.10 & 2.31$\pm$0.15 & 3.10 $\pm$ 0.25   \\
\textbf{PSA-2 (4$\times$64)}&\textbf{30.50} &\textbf{25.35}&\textbf{3.82} & \textbf{0.89$\pm$0.10} & \textbf{1.10$\pm$0.25}   & \textbf{3.25$\pm$0.05}    \\
PSA-3  (8 $\times$ 32)   & 35.05   & 48.10 & 6.68  & 0.90$\pm$0.05 & 2.90$\pm$0.15  & 3.50$\pm$0.15  \\
PSA-4  (8 $\times$ 64)   & 38.56   & 58.80 & 12.86  & 0.95$\pm$0.25  & 3.10$\pm$0.50 & 3.57$\pm$0.30 \\
\bottomrule   
\end{tabular}
\end{table} 
\subsection{Compatibility Analysis of the Proposed PSA-Net with IoT-enabled Smart Devices}
To validate the compatibility of the proposed PSA-Net with IoT-enabled smart devices, we calculate the average inference time in various computational settings. We also report the model sizes and parameters with different cardinalities in Table~\ref{tab:time_complxity}. For this experiment, we tested our PSA-Net on 25 audio samples for prediction. Our analysis ensures that PSA-Net is a computationally efficient approach for multimedia authentication that can be adopted on IoT-enabled voice-controlled devices. For instance, the proposed methods only take $0.89$, $1.10$, and $3.25$ seconds to authenticate an audio clip of 4 seconds on GPU, CPU, and NVIDIA, respectively. Similarly, it can be observed in Table~\ref{tab:time_complxity} that the proposed model PSA-2 with $(4\times64)$ architecture uses very few parameters with a size of $25.35$MB which makes it compatible to run over IoT-enabled smart devices. Although PSA-1 with a $(4\times32)$ configuration performs comparatively slightly better than PSA-2 in compatibility analysis, PSA-2 performs better in voice attack spoofing detection. Lastly, we also used one of the most common matrices called floating point operation per second (FLOPS), a strategy to measure the number of operations required to run a single instance of a deep learning model, to ensure the compatibility of PSA-Net. It can be observed from Table~\ref{tab:time_complxity} that the FLOPS values decrease when the cardinality of the model decreases and vice versa. A similar study conducted in \cite{tu2019deep} also reflects that the model with such a low number of FLOPS runs 3$\times$ faster on edge devices.\vspace{-5pt}

\begin{table}[b]
\caption{ResNets and ResNext models with 18, 34, 50, and 101 layers testing with raw audio, SE and SKIP connections, and the impact of spatial dropout during aggregation testing}\vspace{-5pt}
\label{tab:ablationstudy}
\centering
\begin{tabular}{cc}
\toprule
  Cardinality & EER  \\ \midrule
ResNet-18            & 11.04    \\
SE-ResNet-18             & 6.87   \\
ResNet-34       & 17.54   \\
SE-ResNet-34        & 12.66  \\
ResNet-50         & 38.43  \\
SE-ResNet-50       & 23.54 \\
ResNet-101       & 38.66   \\
SE-ResNet-101        & 30.33 \\
Aggregated Nets-18            & 7.65  \\
SE-Aggregated Nets-18             & 5.50  \\
Aggregated Nets-34       & 8.54  \\
SE-Aggregated Nets-34        & 6.43\\
Aggregated Nets-50         & 30.75 \\
SE-Aggregated Nets-50       & 29.65 \\
Aggregated Nets-101       & 32.76  \\
SE-Aggregated Nets-101        & 29.54 \\
Aggregated Nets-18 (Spatial Dropout)  & 6.40 \\
\textbf{SE-Aggregated Nets-18 (Spatial Dropout)} & \textbf{4.06}   \\
 \bottomrule   
\end{tabular}
 \end{table}

\subsection{Ablation Study and Discussion}
To optimize the PSA-Net architecture and prevent overfitting and underfitting, we investigated the effects of varying channel widths, network densities, and layer topologies. We evaluated the performance of the SE-PSA design by increasing the network's width (number of channels) and density (number of connections between layers). Specifically, we compared configurations similar to the ResNet architecture with 18, 34, 50, and 101 layers, along with the proposed aggregated network with identical layer structures. All networks were trained for 20 epochs on the ASVspoof2019-LA dataset. The results in Table~\ref{tab:ablationstudy} demonstrate that training larger networks (ResNet architectures with 50 and 101 layers) often required significantly more training data and epochs to avoid overfitting to the specific characteristics of the datasets. However, the aggregated networks with SE connections and skip connections achieved superior performance, particularly with 18 and 34 layers. Notably, the 34-layer aggregated network achieved the second-best EER of 8.54\%, while the SE-aggregated network with spatial dropout achieved the lowest EER. Interestingly, the ResNet architecture performed poorly with raw audio samples, resulting in higher EERs. We further evaluated the impact of spatial dropout, a technique that randomly drops out neurons during training to prevent overfitting. Additionally, we compared networks with and without SE connections (which improve feature representation) and skip connections (which allow information flow across layers). As shown in Table~\ref{tab:ablationstudy}, the aggregated network consistently outperformed ResNet architectures, with significant EER improvements. While SE-ResNet variants achieved EERs of 6.87\%, 12.66\%, 23.54\%, and 30.33\%, the corresponding aggregated networks achieved 5.50\%, 6.43\%, 29.65\%, and 29.54\%, respectively. To address potential overfitting concerns due to the increased complexity of the aggregated network, we employed various strategies, like spatial dropout after the aggregation layer. This approach yielded superior results, achieving an EER of 4.06\% when combined with SE and skip connections.

\section{Conclusion}\label{conclusion}
This paper presents a unified spoofing attack detection system, the Parallel Stack Aggregation (PSA) network, designed for direct raw voice processing. Employing a split-transform-merge strategy with multiple cardinal points, it adeptly distinguishes synthetic and replay-based artifacts within speech samples. Through rigorous experimentation and ablation studies, the aggregated residual networks showcased superior performance over simple residual networks when processed using raw voice samples. Moreover, the PSA-Net exhibited an improved ability to capture unique vocal characteristics, minimizing the performance gap between state-of-the-art spoofing attacks compared to methods reliant on handcrafted or spectrogram features employing simple convolutions. The PSA-Net also exhibits the capability to be deployed on less computationally intensive devices. \vspace{-5pt}

\begin{acks}
This study is funded by NSF award number 1815724 and MTRAC ACT award number 292883. The opinions, results, conclusions, or recommendations in this material are solely those of the author(s) and do not necessarily represent NSF or MTRAC ACT views.
\end{acks}

\bibliographystyle{ACM-Reference-Format}
\bibliography{sample-base}

\end{document}